\documentclass[aps,prl,preprint,groupedaddress]{revtex4-1}
\usepackage{graphicx}
\usepackage{dcolumn}
\usepackage{bm}

\begin{document}

\title{Close correlation between superconducting transition temperatures and carrier densities in Te- and S-substituted FeSe thin films}

\author{Fuyuki Nabeshima}
\email[]{cnabeshima@g.ecc.u-tokyo.ac.jp}
\author{Tomoya Ishikawa}%
\author{Naoki Shikama}%
\author{Atsutaka Maeda}%
\affiliation{%
 Department of Basic Science, the University of Tokyo\\
 Meguro, Tokyo 153-8902, Japan
}

\date{\today}

\begin{abstract}
We comparatively investigated the transport properties for S- and Te-substituted FeSe thin films under magnetic fields to clarify the origin of the contrasting behavior of the superconducting transition temperature in S and Te substitution. 
A classical two carrier analysis revealed that the carrier densities of the films increased with increasing Te content, while no significant change was observed for the S-substitution.
This composition dependence of the carrier density well corresponds to the $T_{\mathrm c}$ behavior.
The clear correlation between $T_{\mathrm c}$ and the carrier densities suggests that the structural transition affects the electronic structure in a different manner between Fe(Se,S) and Fe(Se,Te) and that this fact is the direct cause of the difference in the $T_{\mathrm c}$ behaviors at the end point of the structural transition.
\end{abstract}

\pacs{}

\maketitle


Since the discovery of the iron-based superconductors (FeSCs)\cite{Kamihara08}, much research has been devoted to reveal the mechanism of superconductivity in these materials.
An iron chalcogenide superconductor, FeSe\cite{Wu08} is an FeBS with the simplest crystal structure, whose superconducting transition temperature, $T_{\mathrm c}$, is 9 K at ambient pressure.
It shows a structural transition from tetragonal to orthorombic phase at 90 K\cite{PhysRevLett.103.057002}, below which an orbital ordered state was observed\cite{PhysRevB.90.121111,PhysRevLett.113.237001}. 
FeSe exhibits no long-range magnetic order at ambient pressure, while many other iron based superconductors show a magnetic transition at a temperature very close to the structural transition temperature. 
Because the structural transition has a possible electronic origin, it is often called the nematic transition and the interplay between namaticity and superconductivity has received much attention\cite{NPhys.10.97,JPCM.30.023001}. 

A lot of research has been focused on FeSe and S-substituted FeSe because bulk single crystalline samples are available.
With increasing S content, the structural transition temperature decreases, and $T_{\mathrm c}$ once slightly increases and then decrease\cite{PRB.92.121108,PRB.93.224508,PNAS.113.8139,PRB.96.121103}.
Although no significant change in $T_{\mathrm c}$ is observed at the end point of the structural transition, measurements of thermal properties \cite{PNAS.115.1227} and scanning tunneling microscopy/spectroscopy \cite{Hanaguri2018} have revealed an abrupt change in the superconducting gap structure at the end point of the orthorhombic phase, which may suggest the nematic order and its fluctuations have some impact on the superconducting pairing mechanisms. 

Rather different behaviors were observed for another isovalent Te substitution for Se.
Although bulk samples of FeSe$_{1-y}$Te$_y$ with $0.1 \le y \le 0.4$ are not available because of phase separation\cite{Fang08}, we have demonstrated that the single crystalline thin films of Fe(Se,Te) in the whole composition region were grown using pulsed laser deposition\cite{yi15pnas,yiSciRep17}. 
The structural transition temperature is also decreased by Te substitution, and $T_\mathrm c$ is largely enhanced at the end point of the structural transition.
This behavior of $T_\mathrm c$ is in contrast to that of FeSe$_{1-x}$S$_x$, where no such a significant change in $T_\mathrm c$ is observed at the end point of the orthorhombic phase.
The contrasting behavior of $T_{\mathrm c}$ between S and Te substitution suggests that the nematicity has no universal significance on $T_{\mathrm c}$ in these materials\cite{Nabe18.FeSeS}.
Thus, it is very intriguing to elucidate the origin of the difference in the $T_{\mathrm c}$ behaviors between S and Te substitution.

In this letter we report a systematic measurement of Hall effect and magneto-resistance of S- and Te-substituted FeSe thin films. 
A classical two carrier analysis revealed a clear correlation between carrier densities and $T_{\mathrm c}$. 
Our observation suggests that the structural transition affects the electronic structure in a different manner between Fe(Se,S) and Fe(Se,Te).
It also suggests that this fact would be the direct cause of the difference in the $T_{\mathrm c}$ behaviors at the end point of the structural transition.

All the films were grown on LaAlO$_3$ substrates by a pulsed laser deposition method using a KrF laser.
Details of the film growth were described elsewhere\cite{Imai09,Imai10,Nabe18.FeSeS}.
The thicknesses of the grown films were measured by a Dektak 6 M stylus profiler and by X-ray reflectivity measurement.
The electrical resistivity and the Hall resistivity were measured with a standard four-probe method using a physical property measurement system from 2 to 300 K under magnetic fields up to 9 T.

\begin{figure*}[htb]
\includegraphics[width=0.8\linewidth]{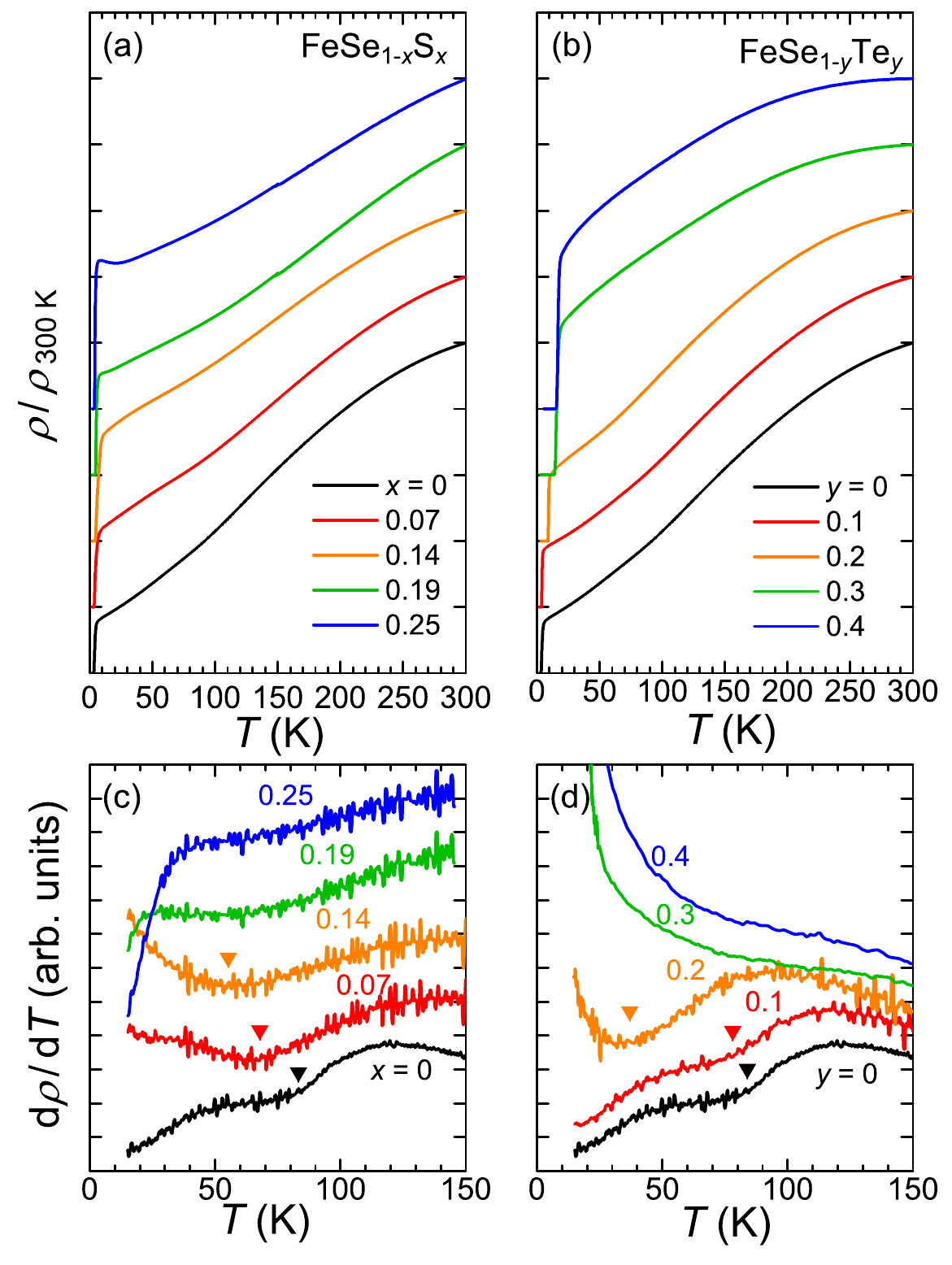}
\caption{Temperature dependence of the dc electrical resistivity of (a) the FeSe$_{1-x}$S$_x$ films and (b) FeSe$_{1-y}$Te$_y$ films. Temperature derivatives of the resistivity of (c) the FeSe$_{1-x}$S$_x$ films and (d) FeSe$_{1-y}$Te$_y$ films. Arrows shows the anomalies due to structural transition.}
\label{G-RT}
\end{figure*}

Figure \ref{G-RT} shows the temperature dependence of the electrical resistivity of the films.
All the films showed the superconducting transition at low temperatures.
The structural transition temperatures, $T_{\mathrm s}$, of the films were determined by the temperature where anomalous behaviors were observed in the temperature derivative of the resistivity (See Fig. \ref{G-RT} (c) and (d)).
$T_{\mathrm s}$ values decreases with increasing substitution amount for both S- and Te-substitution.
We observed no signature of the structural transition for films with $x \ge 0.19$ or $y \ge 0.3$.
FeSe$_{1-y}$Te$_y$ films in the tetragonal phase show very high $T_{\mathrm c}$, while FeSe$_{1-x}$S$_x$ has small $T_{\mathrm c}$ values even in the tetragonal phase.
Note that a slight up-turn behavior was observed at low temperatures in the $R$-vs-$T$ curve of the $x=0.25$ film. 
We observed an up-turn behavior with a clear kink in the $R$-vs-$T$ curve in other films with similar compositions\cite{Nabe18.FeSeS}, which could be attributed to a magnetic transition because the $R$-vs-$T$ behavior is very similar to what is observed at the antiferromagnetic transition in FeSe under hydrostatic pressure\cite{NatCom.7.12146}.
Although the up-turn in the $R$-vs-$T$ curve observed in this study is weaker, this would have the same origin.

In our previous paper, we reported that FeSe$_{1-y}$Te$_y$ films showed $T_{\mathrm s}$ of approximately 45 K for $y = 0.3$, while the film with $y = 0.3$ in this study showed no structural transition. 
This could be attributed to the difference in the strength of the lattice strain between films in the present and previous study; the film in this study has shorter $a$-axis length ($a \sim 3.76$ \AA) than films in the previous study ($a = 3.77-3.78$ \AA)\cite{yiSciRep17}. 
This is consistent with the fact that the structural transition temperature $T_{\mathrm s}$ of FeSe decreases when the strain become more compressive\cite{Nabe18.FeSeStrain}.

\begin{figure*}[htb]
\includegraphics[width=\linewidth]{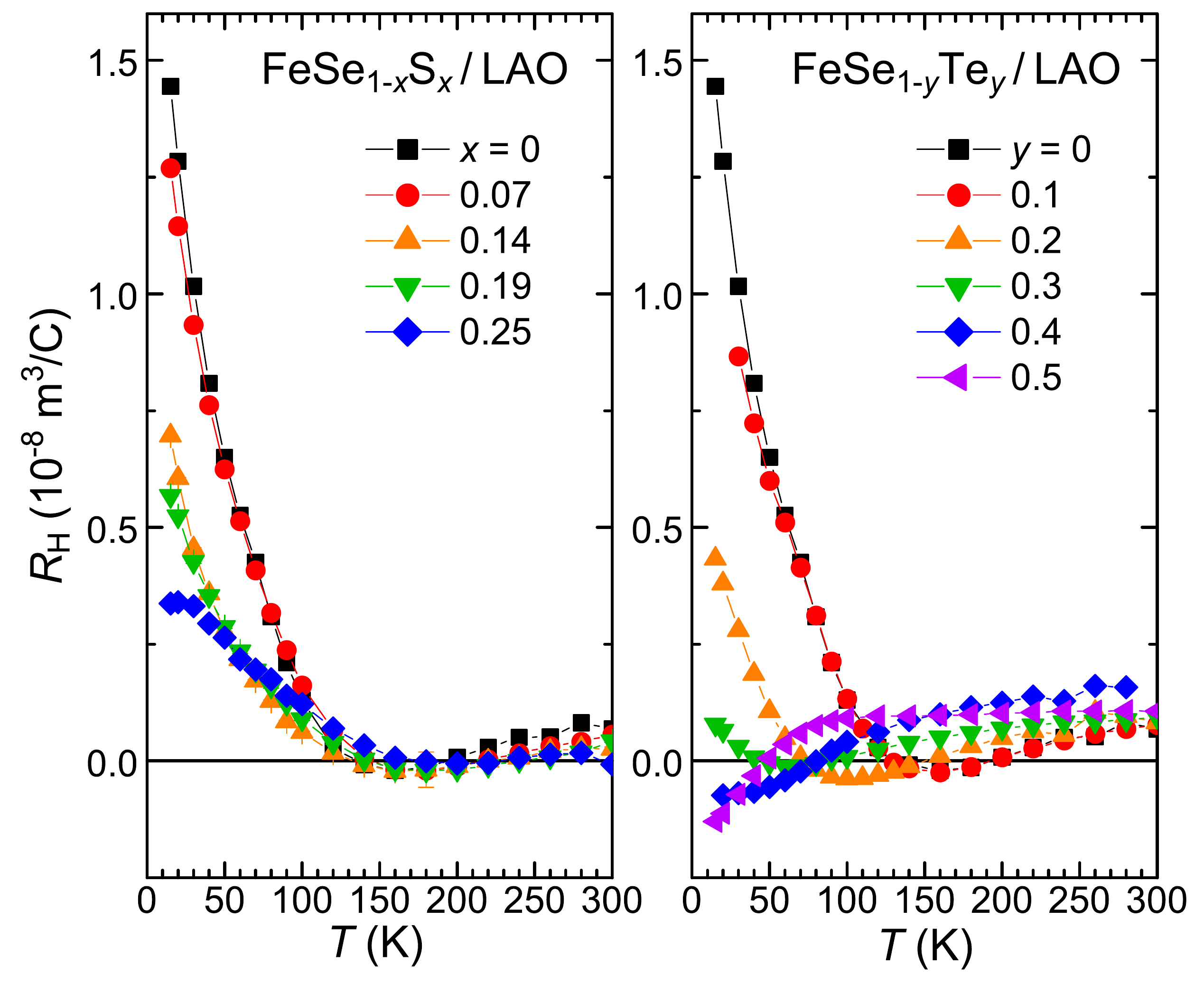}
\caption{Temperature dependence of the Hall coefficients, $R_\mathrm H$, of (a) the FeSe$_{1-x}$S$_x$ films and (b) FeSe$_{1-y}$Te$_y$ films.}
\label{G-Hall}
\end{figure*}

Figure \ref{G-Hall} shows the temperature dependence of the Hall coefficients, $R_\mathrm H$, of the FeSe$_{1-x}$S$_x$ and FeSe$_{1-y}$Te$_y$ films.
$R_\mathrm H$ of the FeSe film largely increased below $T < 100$ K. 
The enhancement of $R_\mathrm H$ at low temperatures was decreased by both S and Te substitution.
For Te substitution, $R_\mathrm H$ at low temperatures become very small after the structural transition disappeared.
This behavior is very similar to those of FeSe$_{1-y}$Te$_y$ films on CaF$_2$ substrates\cite{ys16jpsj}.
On the other hand, no significant change in $R_\mathrm H$ was observed between the films with and without the structural transition in the case of S substitution.
$R_\mathrm H$ at low temperatures was large even after the structural transition disappeared for S substitution.
The $R_\mathrm H$-vs-$T$ behaviors of the S-substituted films are similar to those of bulk samples\cite{PhysRevB.93.104502,JLTP.185.467}, while that of the FeSe film is different from that of bulk FeSe\cite{PNAS.111.16309,PhysRevB.90.144516}.
No clear signature of the possible AFM transition was observed in $R_H$.

In order to investigate the nature of charge carriers in the films we also performed magnetoresistance measurements.
Both electron- and hole-type carriers contribute to the electric conduction in a multiband system like iron chalcogenides.
We considered one electron band and one hole band representing the multiple bands and applied the text-book approach for multiband materials. 
In a classical two-carrier model, the resistivity tensor is expressed as
\begin{eqnarray}
\rho_{xx} (0) &= \frac{1}{e(n_{\mathrm h}\mu _{\mathrm h}+n_{\mathrm e}\mu _{\mathrm e})},\\
\frac{\rho_{xx} (B) - \rho_{xx} (0)}{\rho_{xx} (0)} &= \frac{n_{\mathrm h}n_{\mathrm e}\mu_{\mathrm h}\mu_{\mathrm e}(\mu_{\mathrm h}+\mu_{\mathrm e})^2}{(n_{\mathrm h}\mu _{\mathrm h}+n_{\mathrm e}\mu _{\mathrm e})^2} B^2, \\
\rho_{yx}(B) &= \frac{n_{\mathrm h}\mu_{\mathrm h}^2-n_{\mathrm e}\mu_{\mathrm e}^2}{e(n_{\mathrm h}\mu _{\mathrm h}+n_{\mathrm e}\mu _{\mathrm e})^2} B,
\end{eqnarray}
where $n_{\mathrm h}$, $n_{\mathrm e}$, $\mu_{\mathrm h}$, and $\mu_{\mathrm e}$ are the hole density, the electron density, the hole mobility, and the electron mobility, respectively.
Validity of application of the two-carrier model to iron chalcogenides has been confirmed by a recent THz Faraday rotation measurement, where the two-carrier model perfectly reproduced the measured $\sigma _{xx} (\omega) $ and $\sigma_{yx} (\omega)$\cite{arXiv:1903.00897}.
We evaluated the carrier densities and mobilities of the films with the measured data of $R_{\mathrm H}$ and the magnetoresistance, assuming $n_{\mathrm h} = n_{\mathrm e}$.

\begin{figure*}[htb]
\includegraphics[width=\linewidth]{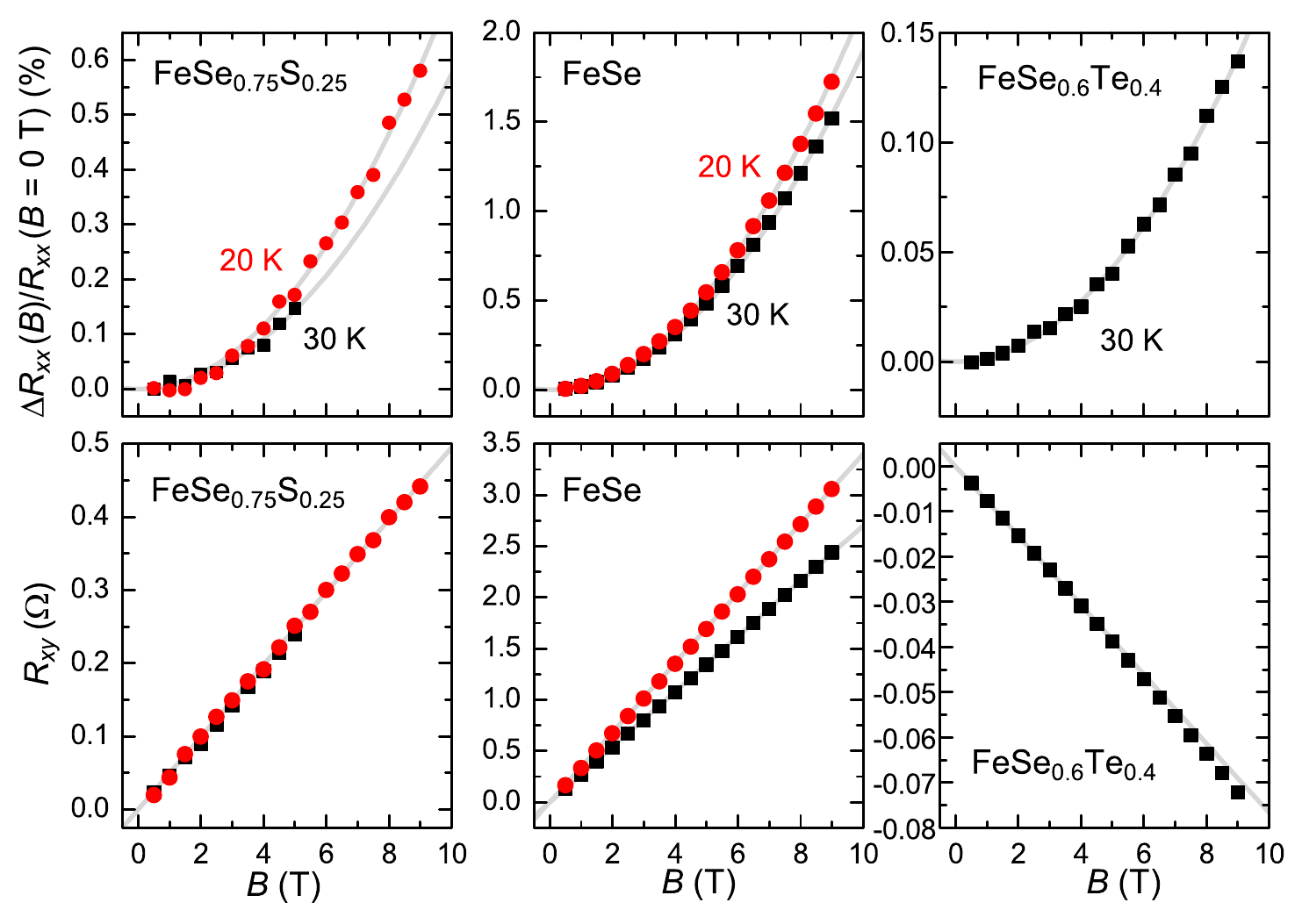}
\caption{Magneto-resistance, $(\rho (B) - \rho(0\ \mathrm T)) / \rho(0\ \mathrm T)$, (upper panels) and Hall resistance, $R_{xy} (B)$, (lower panels) as a function of the magnetic field of FeSe(center), $x=0.25$ (left), and $y=0.4$ (right) films at 20 K (red circle) and 30 K (black square). The Gray lines are the fitted curves for these data by the two-carrier model.}
\label{G-MRHall}
\end{figure*}

Figure \ref{G-MRHall} shows the magneto-resistance, $(\rho (B) - \rho(0\ \mathrm T)) / \rho (0\ \mathrm T)$, and Hall resistance, $R_{xy} (B)$ as a function of the magnetic field of FeSe, $x=0.25$, and $y=0.4$ films.
The magneto-resistance and Hall resistance are well fitted by $B^2$ and $B$-linear curves, respectively, for all samples.
Negative magneto-resistance due to weak localization was not observed, which may support the supposition that the up-turn behavior in the $R$-vs-$T$ curve observed for the $y=0.25$ film is caused by the magnetic transition.

\begin{figure*}[htb]
\includegraphics[width=0.75\linewidth]{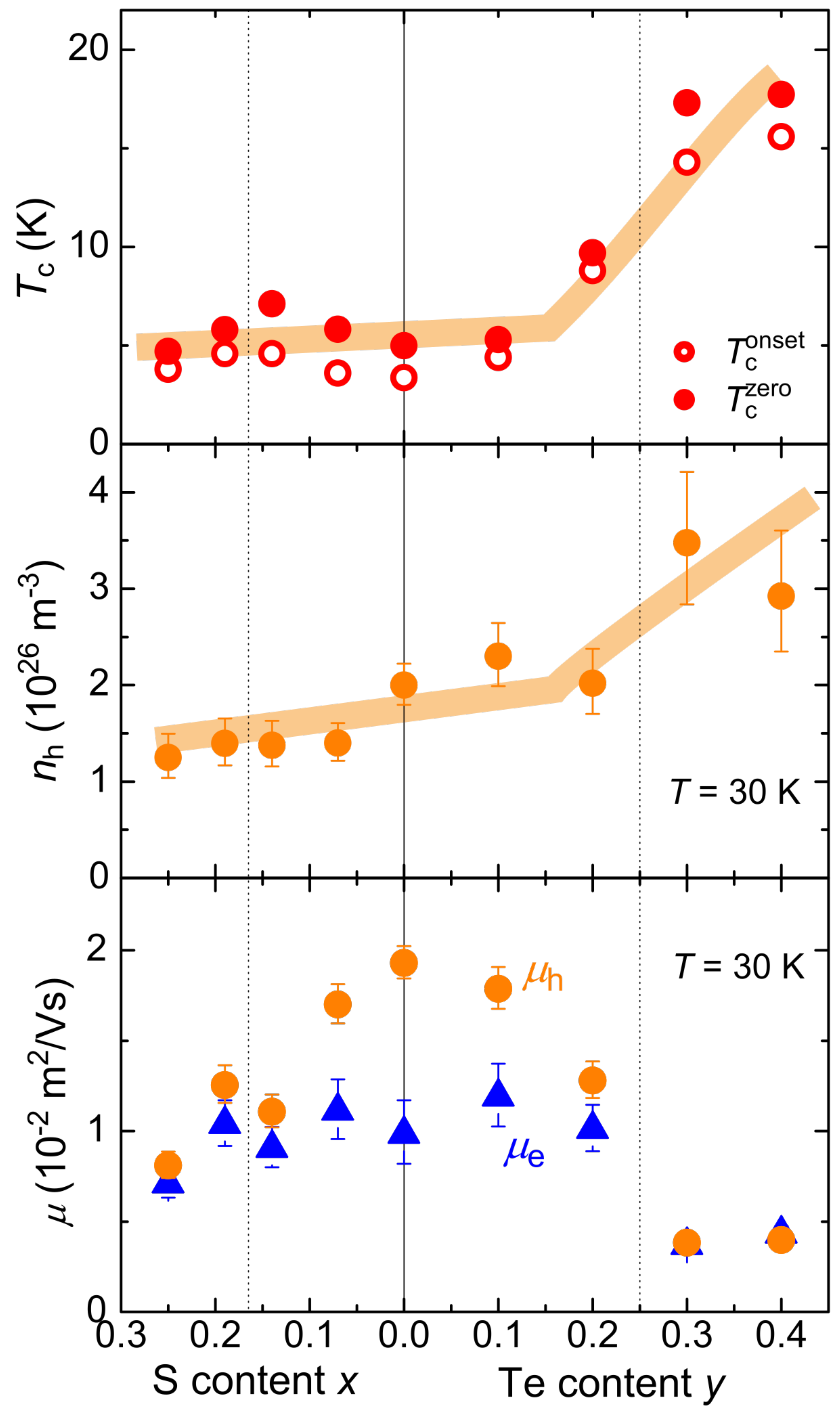}
\caption{Results of the analysis with the two carrier model. The composition dependence of (a) the superconducting transition temperature, $T_\mathrm c$, (b) the carrier densities, $n_\mathrm h$ and $n_\mathrm e$ at 30 K, and (c) the mobilities, $\mu_\mathrm h$ and $\mu_\mathrm e$ at 30 K. }
\label{G-CarrierDensity}
\end{figure*}

Figure \ref{G-CarrierDensity} sbows the results of the analysis with the two carrier model as a function of the composition of the films.
The error bars in the carrier densities and mobilities shows deviations when the assumed ratio of the carrier densities were varied in the range of $0.4 < n_{\mathrm h} / (n_{\mathrm h} + n_{\mathrm e} )< 0.6$ in the analysis.
For S substitution, the carrier densities were almost constant, independent on the S content, while the hole and electron mobilities decreased with increasing substituting amount.
An angle-resolved photoemission spectroscopy (ARPES) measurement for bulk FeSe$_{1-x}$S$_x$ revealed that the Fermi surface becomes large for samples in the tetragonal phase\cite{PRB.96.121103}. 
Such a slight increase in carrier densities was not observed in the present study, which would be due to our assumption that the ratio of the carrier densities is constant for all the films in the analysis. 
For Te substitution, on the other hand, the carrier densities largely increased with increasing Te content; the $y=0.4$ film have $n_{\mathrm h}$ more than 1.5 times lager than that of the FeSe film. %
The mobilities decreased with increasing Te content, as the same as S substitution.
Thus, we observed a contrasting behavior of the carrier densities between S and Te substitution; the carrier densities increased with increasing Te content, while no significant change was observed for S substitution.
This composition dependence of the carrier densities well corresponds to the change of the $T_{\mathrm c}$ values. 
Therefore, this result indicates a close correlation between the carrier densities and $T_{\mathrm c}$ in our films.

We should note that the error bars of $n_{\mathrm h}$ for the $y =$0.3 and 0.4 films are rather large and the $n_{\mathrm h}$ values to be obtained would become as small as that of the FeSe film when the assumed value of $n_{\mathrm h} / (n_{\mathrm h} + n_{\mathrm e})$ is changed.
In this case, however, the $n_{\mathrm e}$ values become larger, and thus, the above conclusion does not change.

The similar correlation between $T_{\mathrm c}$ and carrier density was observed in FeSe films under various degrees of lattice strain\cite{Nabe18.FeSeStrain,arXiv:1807.01273}.
These results suggest the close relation between $T_{\mathrm c}$ and the carrier densities in iron chalcogenides. 
We already mentioned, based on our results, that the electronic nematicity does not have a universal significance in superconductivity in iron chalcogenides. 
The difference of the behavior of the carrier densities between Te and S substitution suggests that the structural transition affects the electronic structure differently between FeSe$_{1-y}$Te$_y$ and FeSe$_{1-x}$S$_x$. 
This is the direct origin of the difference of $T_{\mathrm c}$ behavior at the end point of the orthorhombic phase, manifested in the close correlation between $T_{\mathrm c}$ and the carrier densities.

In conclusion, we comparatively investigated transport properties of FeSe$_{1-x}$S$_x$ and FeSe$_{1-y}$Te$_y$ thin films under magnetic fields. 
Enhancement of the Hall coefficient, $R_H$, of FeSe films at low temperatures was reduced by both S and Te substitution.
Fe(Se,Te) showed small $R_H$ values near zero at low temperatures in the tetragonal phase, while $R_H$ values remained rather large for Fe(Se,S). 
The analysis of the magneto-resistance and the Hall resistance using a classical two carrier model revealed that the carrier densities of the films increased with increasing Te content, while no significant change was observed for the S-substitution.
This suggests that the structural transition affects the electronic structure differently between Fe(Se,S) and Fe(Se,Te).
The clear correlation between $T_{\mathrm c}$ and the carrier densities indicates this fact is the direct cause of the difference in the $T_{\mathrm c}$ behaviors at the end point of the structural transition.

\begin{acknowledgments}
We would like to thank K. Ueno for the X-ray diffraction measurements of the films.
We also thank M. Hanawa for the thickness measurements of the films.
This research was supported by the Murata Science Foundation and JSPS KAKENHI Grant Numbers 18H04212 and 19K14651. 
\end{acknowledgments}

\clearpage

%


\end{document}